# ANIMATING FERMI – A COLLABORATION BETWEEN ART STUDENTS AND ASTRONOMERS

Laurence Arcadias, Animation Department, Maryland Institute College of Art, Baltimore, MD 21217, U.S.A. E-mail: <larcadias@mica.edu>

Robin Corbet, CRESST, University of Maryland, Baltimore County, Baltimore, MD 21250, U.S.A.; X-ray Astrophysics Laboratory, NASA Goddard Space Flight Center, Greenbelt, MD 20771, U.S.A. E-mail: <corbet@umbc.edu>

**Abstract**

**Undergraduate animation students at the Maryland Institute College of Art teamed up with scientists from the Fermi Gamma-ray Space Telescope to produce a set of animations on several astronomy topics. We describe the process and discuss the results, including educational benefits and the cross-cultural experience. These animations are freely available online.**

## Scientific Background

The Fermi Gamma-ray Space Telescope is a large international project. Named after the physicist Enrico Fermi (1901–1954), the satellite carries two types of detectors designed to study gamma rays from a variety of astronomical sources [1,2]. Fermi has been successfully operating since 2008, and it has enabled many discoveries in the field of high-energy astrophysics for a variety of objects whose distances range from the relatively nearby, such as the Sun and the Moon, to distant galaxies containing extremely massive black holes [3]. The production of gamma rays requires extreme mechanisms to create this most energetic form of "light". In the U.S.A., Fermi is jointly supported by NASA and the Department of Energy, with a number of Fermi scientists working at the NASA Goddard Space Flight Center (GSFC). Corbet is the operations lead in the Fermi Science Support Center at GSFC

## Artistic Background and Project

Maryland Institute College of Art (MICA) is a leading visual arts school offering undergraduate and graduate programs. Students in the Animation Department are trained to use digital and traditional 2D/3D tools as best suited to expressing their creative visions. The advanced 2D junior-level class taught by Arcadias gives students a chance to think about animation in broad ways including animated documentaries that led us to investigate art and astronomy in a similar vein to "Dance your PhD" or Björk's Biophilia [4].

In spring 2014, following discussions between LA and RC we initiated a collaboration between MICA and Fermi team members at GSFC. The goal was to have the 17 students in the class create animations based on Fermi research with the scientists mentoring their work. Students were invited to use scientific concepts as a source of inspiration rather than producing direct visualizations while still remaining true to the scientific ideas.

## Process

*Step one:* Corbet came to MICA and explained some basic astrophysical concepts and Fermi's fields of research. Students formed small groups and spent a week on storyboarding.

*Step two:* Students visited NASA GSFC. After a guided tour of several areas, including the James Webb Space Telescope clean room, the students pitched their animation concepts to the scientists and scientist mentors were identified.

*Step three*: The students spent five weeks producing five short animations based on the topics described below, all compiled under the general title of *Animating Fermi*. A Tumblr page was created as a way to interact with the scientist mentors who could leave feedback as the work was posted.

*Step four*: During a second trip to NASA, a screening of the animations was hosted at the Goddard Visitor Center. Comments from the lively discussion that followed were incorporated into the final versions. A short documentary of the experience can be seen here: <https://vimeo.com/91361066>

## Science/Animation Topics

Five topics were chosen based on the scientists' expertise and students' interests:

*Fermi bubbles:* Large gamma-ray emitting regions extending above and below the center of our Milky Way galaxy were discovered with Fermi in 2010. The origin of these "bubbles" is unclear, but a possible connection with the black hole at the center of our Galaxy is suspected.

*Binary stars:* Stars are often found in binary systems: a pair of stars orbiting a common center of mass. If one component is a black hole or neutron star, interactions may produce gamma ray emission that varies periodically with the orbital phase.

*Space debris:* There is an increasing amount of "junk" in orbit around the Earth. This ranges in size from dust to entire satellites. The Fermi satellite is equipped with a rocket motor, one purpose of which is to enable Fermi to maneuver to avoid colliding with large objects. Such a maneuver was successfully performed in 2012 when a collision with a defunct cold-war era spy satellite was averted.

*Dark matter:* It is now known that most of the matter in the Universe does not emit light and does not consist of already known forms of matter. Several speculative ideas have been proposed for its nature. Theoretical models generally predict a high density of dark matter at the center of the Milky Way, perhaps resulting in the production of gamma rays.

*Cosmic rays:* Fermi is also sensitive to cosmic rays, which are high-energy particles that mainly come from outside the Solar System. They are regarded as either signal or noise depending on the scientific project. Cosmic rays were initially thought to originate from the Earth, until a balloon flight by Victor Hess in 1919 showed that they do come from space.

**Fig. 1. Students (and scientist) at work. (© Laurence Arcadias.)**

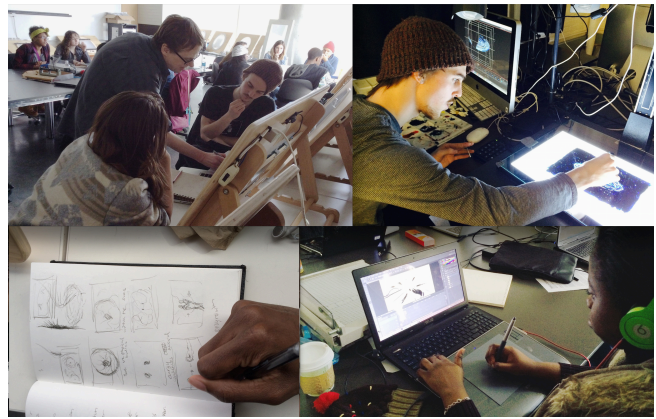

## Final Animations

*Animating Fermi* is available under a creative commons license at: <https://vimeo.com/94023644>. *Animating Fermi* has

been presented at several venues and conferences: GSFC Visitor Center (April 2014), VERITAS Workshop, New York (May 2014), Be X-ray Binary meeting, Valencia, Spain (July 2014), Chicago American Astronomical Society (September 2014; poster), SIGGRAPH 14: 3rd Annual Faculty Submitted Student Work Exhibit, Art and Algorithms digital art festival in Florida (October 2014). The animations have the potential to be used in a variety of education and public outreach ways. Several educators reported that they intended to show them in their university-level science and/or art classes.

### Educational Outcomes

Many MICA students were already interested in science, so they appreciated the opportunity to collaborate on a project with astronomers. The visit to various NASA facilities gave them a better understanding of how scientists operate in such environments. In the class evaluations students commented on how approachable scientists were and how they expressed interest in their animations. A few challenges were identified during the production phase, such as how far the students could go in their interpretation of a concept and stay scientifically accurate without being too literal. Finding the right metaphors to explore a concept visually was one of the keys to success. A student remarked how scientists themselves used metaphors when explaining a concept. For an examination of metaphor use in science see e.g. [5]. Making the animations understandable to a broad audience and still be appealing to an expert was also addressed. The final critique with the scientists gave the students a taste of scientific thinking and motivated them to explore this further in their own approach.

### Bridging Scientific and Artistic Cultures

While the nominal division between science and the humanities has frequently been debated [6,7], there are some current initiatives to incorporate art into "STEM" disciplines as "STEAM" [e.g. 8]. Our collaboration illustrates how artists can be involved in science in different ways, either using art/animation for educational or illustrative purposes or exploring a scientific concept to help us perceive an idea in an unexpected way. Some of the students expressed surprise at seeing scientists debating facts, perhaps expressing a misconception of science as simply a body of knowledge rather than primarily a process. One student, referring to the beauty of many NASA released images, stated that it wasn't possible to "compete" with these. We note that while scientific images are directly derived from data, those chosen for public release have often been picked for their inherent visual appeal, and aesthetic choices have been made in the conversion between numerical data and visual appearance [9].

Our project revealed some commonalities between artistic and scientific approaches [10]. One of the students commented that both artists and scientists have similarities in working using creativity and intuition to deal with their particular material. Scientists also reacted positively to the expression of their work in new and playful ways.

Conversely, an NSF report on the public image of the science community [11] noted that scientists and engineers are typically portrayed as "unattractive reclusive socially inept white men or foreigners working in dull unglamorous careers". In two animations scientists were directly portrayed. One was a historical figure (Hess) giving less scope for interpretation. In the other case (dark matter), the scientific figure was an older white male, but in that case, it was also a humorous way of talking about clichés and how to question ideas.

### Further Development

Continuing this collaboration, an internship was created at GSFC in summer 2014: Turner Gillespie (MICA) and Sylvia Zhu (Fermi) produced a visualization and sonification of gamma-ray bursts ("GRB Suite" [12]). We are continuing the MICA/Fermi project in 2015 and are investigating other ways to expand these art/astronomy interactions.

We thank the MICA student animators and Fermi scientists; C. Fluke, R. Kingsburgh and J. Perez-Gallego for useful comments; the MICA Office of Community Engagement; and CRESST/UMBC for sponsoring the summer internship.

**Fig. 2. Top: Students during final critique at NASA. (© L. Arcadias. Photo: E. Bershof.). Bottom: Space debris animation still.**

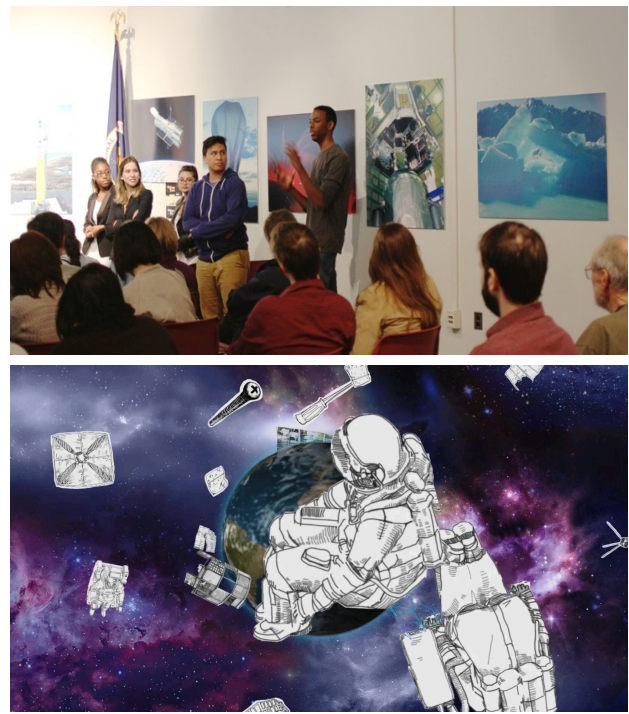